\documentclass[twocolumn,prb, aps,showpacs, superscriptaddress]{revtex4-1}
\usepackage{amsmath}
\usepackage{graphicx, epstopdf}
\usepackage{amsfonts}
\usepackage{amssymb}
\usepackage{dcolumn}% Align table columns on decimal point
\usepackage{bm}% bold math
\usepackage{subfigure}
\usepackage{mathptmx}

\begin{document}

\title{Band gap anomaly and topological properties in lead chalcogenides}

\author{Simin Nie}
\author{Xiao Yan Xu}
\affiliation{Beijing National Laboratory for Condensed Matter Physics,
  and Institute of Physics, Chinese Academy of Sciences, Beijing
  100190, China}

\author{Gang Xu}
 \email{gangx@iphy.ac.cn}
\affiliation{Beijing National Laboratory for Condensed Matter Physics,
  and Institute of Physics, Chinese Academy of Sciences, Beijing
  100190, China}

\author{Zhong Fang}
 \affiliation{Beijing National Laboratory for Condensed Matter Physics,
  and Institute of Physics, Chinese Academy of Sciences, Beijing
  100190, China}
\affiliation{Collaborative Innovation Center of Quantum Matter,
  Beijing, 100084, China}

\date{\today}

\begin{abstract}
   Band gap anomaly is a well-known issue in lead chalcogenides PbX (X=S, Se, Te, Po).
   Combining $ab~initio$ calculations and tight-binding (TB) method, we have studied the
   band evolution in PbX, and found that the band gap anomaly in PbTe is mainly related to
   the high onsite energy of Te 5$s$ orbital and the large $s$-$p$ hopping originated from
   the irregular extended distribution of Te 5s electrons. Furthermore, our calculations
   show that PbPo is an indirect band gap (6.5 meV) semiconductor with band inversion at
   L point, which clearly indicates that PbPo is a topological crystalline insulator (TCI).
   The calculated mirror Chern number and surface states double confirm this conclusion.
\end{abstract}
\maketitle

\section{Introduction}

Rock-salt chalcogenides represent a significant group of functional materials, in which
many novel properties are discovered, including superconductivity~\cite{matsushita2006type},
thermoelectricity~\cite{wood1988,wang2011heavily}, ferroelectricity~\cite{lebedev1994},
optoelectronics~\cite{Akimov,liu2010} and spintronics~\cite{Jin2009,grabecki2007}. Recently,
this class of materials has attracted considerable attention again due to the realization of
the nontrivial topological properties, so called TCI~\cite{Fu2011,weng2014topological} , in
SnTe~\cite{hsieh2012,tanaka2012}. In contrast to the widely studied $Z_2$ topological insulator
(TI)~\cite{hasan2010colloquium,qi2011topological,weng2014exploration,weng2014transition,
weng2015large,xu2011chern,xu2014quantum,niequantum}, Dirac cones lying on the TCI surfaces are
protected by the mirror symmetry, rather than time reversal (TR) symmetry. Therefore, TCI phase
is associated with a new topological invariant called mirror Chern number, which classifies and
distinguishes TCI from TI and an ordinary insulator. Importantly, a quantized $\pi$-Berry phase
induced by the band inversion is needed and essential in TCI to characterize its topological
properties such as Dirac cones.

Though the discovery of TCI enriches people's understanding of the topological classification,
few TCIs are realized in experiment. In order to satisfy the requirement of the potential
device applications, it is urgent to find more TCIs, especially those with large band gap and
high working temperature. As we all know, most lead chalcogenides adopt the same structure as
SnTe, where the band inversion at L point is driven by the spin-orbit coupling (SOC). Considering
that lead is much heavier than tin, it is natural to postulate that PbX may be large band
gap TCI due to the much stronger SOC. However, until now, no TCI phase is realized in PbX.
Therefore, it is important to figure out the main factors of the band gap evolution in rock-salt
chalcogenides with the purpose of achieving large band gap TCIs. Based on previous studies,
there is a famous empirical relation~\cite{moss1959optical} between band gap $E_G$ and lattice
constant $a_0$, which states that $E_G = E(L_6^-)-E(L_6^+)$ has a linear relationship
with $1/a_0^2$ for a series of relevant rock-salt semiconductors. However, this relation
is invalid in lead chalcogenides, where PbTe exhibits a well-known anomaly that
$E_G(\text{PbS})>E_G(\text{PbTe})>E_G(\text{PbSe})$~\cite{Donald1962,dalven1971}, even
though the lattice constant $a_0$(PbTe) is much larger than $a_0$(PbSe). This band gap
anomaly has been a long-standing question and remains under debate.

\begin{figure}[tbp]
\includegraphics[clip, width=3.5in]{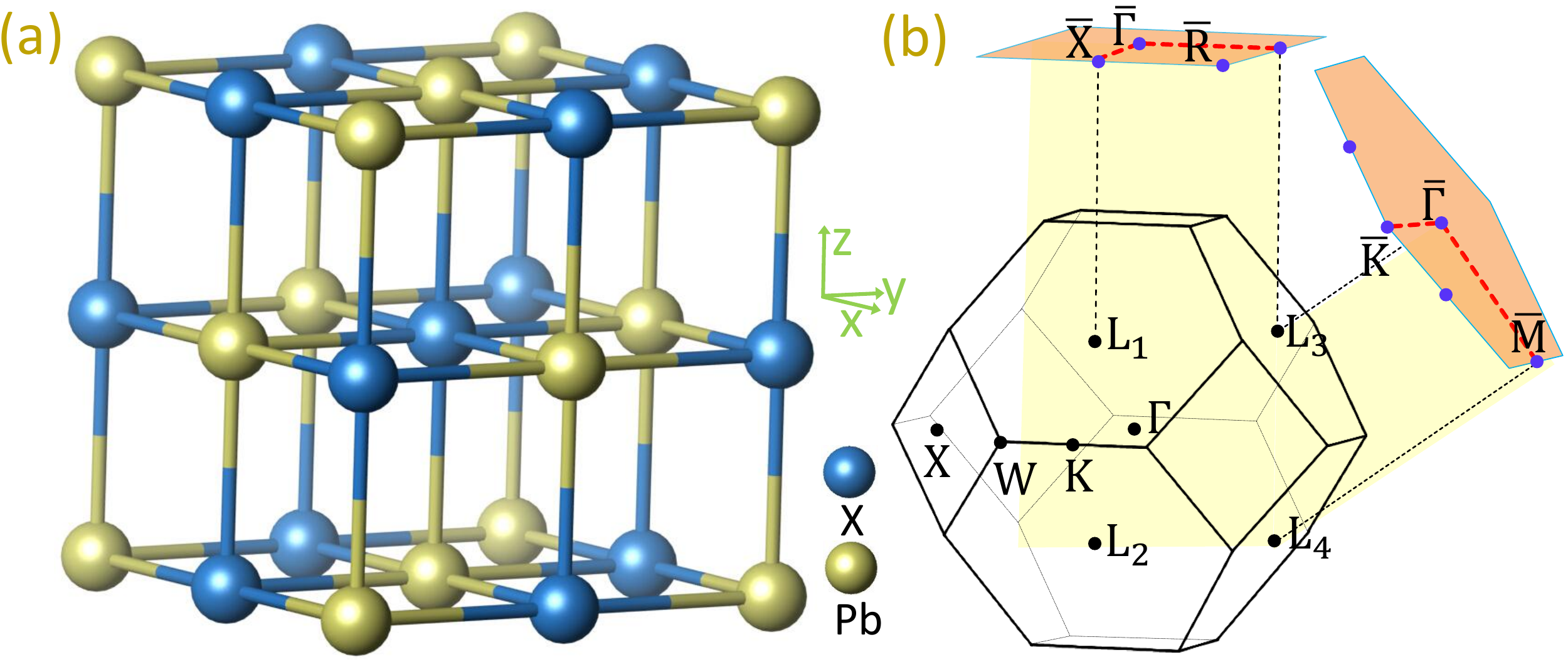}
%\captionsetup{justification=justified}
      \caption{(color online) (a) The crystal structure of PbX with space group $Fm\overline{3}m$.
               (b) First Brillouin zone (BZ) of bulk and the projected surface BZ on (001) and (111)
               planes. There are four L points (L$_1$, L$_2$, L$_3$, L$_4$) in
               the bulk first BZ. For the (001) face, L$_1$ and L$_2$ are projected
               to the $\overline{\text{X}}$, while  L$_3$ and L$_4$ are projected to another $\overline{\text{X}}$.
               For the (111) face, L$_3$ is projected to $\overline{\Gamma}$, while L$_1$,
               L$_2$ and L$_4$ are projected to $\overline{\text{M}}$.
               The light yellow region is invariant under the mirror symmetry $\hat{m}_{(1\overline{1}0)}$.}
\label{structure}
\end{figure}

In this paper, by combining $ab~initio$ calculations and TB method, we have studied the band
evolution and band gap anomaly in PbX. Our studies show that the band gap anomaly in PbTe is
related to the delocalized 5$s$ electrons of Te tightly. Comparing to S and Se, the $5s$
electrons of Te show much low binding energy (high onsite energy) and very extended distribution
caused by the huge screening effect of the numerous of interior electrons. As a result, Te $5s$
orbital pushes up the Pb $p$ orbital through the huge $s$-$p$ hybridization, leading to a large
band gap for PbTe. Moreover, our calculations show that $E_G(\text{PbPo})$ roughly agrees with
the linear relation $E_G\propto$1/$a_0^2$ formed by PbS and PbSe, and it becomes a negative number,
which means that band inversion happens at L point in PbPo. Our detailed Non-local Heyd-Scuseria-
Ernzerhof (HSE) hybrid functional calculations show that PbPo is an indirect band gap semiconductor
with a quantized $\pi$-Berry phase, which clearly indicates that PbPo is a TCI. The calculated mirror
Chern number and surface states double confirm this conclusion.

This paper is arranged as follows. In section II we will introduce the details of
the $ab~ initio$ calculations and TB model. In section III, based on the TB model,
we will study the origin of the band gap anomaly. In section IV we will focus on the
electronic structure and topological properties of PbPo. Finally, section V contains
a summary of this work.

\section{\emph{Ab inito} CALCULATIONS and TB method}

Our $ab ~initio$ calculations are carried out by the projector augmented wave (PAW)
method~\cite{blochl1994,kresse1999} implemented in Vienna $ab~ initio$ simulation
package (VASP)~\cite{kresse1996_1,kresse1996_2}. Experimental lattice constants, with
$a_0$ = 5.942~\cite{mehmood2010}, 6.124~\cite{aliev2008}, 6.460~\cite{minikayev2011},
and 6.59 \AA~\cite{witteman1960} for PbS, PbSe, PbTe, and PbPo are adopted in our
calculations. The exchange and correlation potential is treated within the generalized
gradient approximation (GGA) of Perdew-Burke-Ernzerhof type~\cite{Perdew1996}. Considering
the possible underestimation of the band gap by GGA, HSE hybrid functional~\cite{heyd2003hybrid}
is further supplemented to improve the accuracy of the band gap. The cutoff energy of
the plane wave expansion is 500 eV, and 11 $\times$ 11 $\times$ 11 k-point grids are used
in the self-consistent calculations. SOC is consistently considered in the calculations.
Modified Becke-Johnson (mBJ)~\cite{tran2009accurate} calculations are performed using the
all-electron full-potential linearized augmented plane-wave (FP-LAPW) method implemented in
the WIEN2k package~\cite{blaha2002wien2k}.

In order to study the band evolution and the band gap anomaly accurately, a general eight-band
Slater-Koster~\cite{slater1954simplified} TB model with bases
\{$s^{C}$, $p^{C}_x$, $p^{C}_y$, $p^{C}_z$, $s^{A}$, $p^{A}_x$, $p^{A}_y$, $p^{A}_z$\}
are constructed, in which, besides the $p$-$p$ hoppings, the $s$-$p$ hybridizations are
taken into account too:

\begin{align}
H_0=\sum_{\mu\nu}\sum_{\alpha\beta}\sum_{ij}(t_{\mu\nu,ij}^{\alpha\beta}+
\epsilon_{\mu}^{\alpha}\delta_{\mu\nu}\delta_{\alpha\beta}\delta_{ij})c_{\mu \alpha}^+(i)c_{\nu \beta}(j)
\end{align}%}
where $\mu,\nu$ = cation, anion label the sublattices. $\alpha, \beta$ label the $s$, $p_x$,
$p_y$ and $p_z$ orbitals. $i,j$ label the atomic sites. $t_{\mu\nu,ij}^{\alpha\beta}$ represents the
corresponding hopping parameters. $\epsilon_{\mu}^{\alpha}$ means the onsite energy of $\alpha$
orbital on $\mu$ sublattice. $c_{\mu \alpha}^+(i)$ ($c_{\nu \beta}(j)$) creates (annihilates)
an $\alpha$ ($\beta$) electron on $\mu$ ($\nu$) sublattice at site $i$ ($j$). The interactions
up to the fourth-nearest neighbors have been considered in this paper, which leads to totally 32
independent parameters as listed in Table \ref{fittedparameters}. We note that all parameters in
Table \ref{fittedparameters} are reliable and very close to previous study\cite{lach2000}.
In momentum space, the Hamiltonian is given by,
\begin{equation}
H_0=\sum_{\mathbf{k}}\Psi^{+}_\mathbf{k}~H_0(\mathbf{k})~\Psi_\mathbf{k}
\end{equation}
with
{\small
\begin{eqnarray}\label{H00}
H_0(\mathbf{k})  &=&\begin{pmatrix}
       H_0^{CC}&    H_0^{CA}\\
       H_0^{AC}&    H_0^{AA}\\
      \end{pmatrix} \nonumber\\
      &=&\begin{pmatrix}
	     H_{1,1}&      G_{1,2}&      G_{1,3}&      G_{1,4}&    H_{1,5}&  G_{1,6}&    G_{1,7}&    G_{1,8}\\
		 G_{2,1}&      H_{2,2}&      H_{2,3}&      H_{2,4}&    G_{2,5}&  H_{2,6}&    H_{2,7}&    H_{2,8}\\
         G_{3,1}&      H_{3,2}&      H_{3,3}&      H_{3,4}&    G_{3,5}&  H_{3,6}&    H_{3,7}&    H_{3,8}\\
         G_{4,1}&      H_{4,2}&      H_{4,3}&      H_{4,4}&    G_{4,5}&  H_{4,6}&    H_{4,7}&    H_{4,8}\\
         H_{5,1}&      G_{5,2}&      G_{5,3}&      G_{5,4}&    H_{5,5}&  G_{5,6}&    G_{5,7}&    G_{5,8}\\
         G_{6,1}&      H_{6,2}&      H_{6,3}&      H_{6,4}&    G_{6,5}&  H_{6,6}&    H_{6,7}&    H_{6,8}\\
         G_{7,1}&      H_{7,2}&      H_{7,3}&      H_{7,4}&    G_{7,5}&  H_{7,6}&    H_{7,7}&    H_{7,8}\\
         G_{8,1}&      H_{8,2}&      H_{8,3}&      H_{8,4}&    G_{8,5}&  H_{8,6}&    H_{8,7}&    H_{8,8}\\
\end{pmatrix}%
\end{eqnarray}}where $H_{i,j}$ and $G_{i,j}$ mean the matrix elements are real and imaginary,
respectively. The detailed descriptions of all matrix elements are given in the Appendix.

When SOC is taken into account, the Hamiltonian size will be doubled. The final
Hamiltonian with SOC can be written in the form,
\begin{widetext}
\begin{align}
 \label{HH}
H(\mathbf{k})  &= H_0(\mathbf{k})+H_{so}(\mathbf{k})
=\begin{pmatrix}
    H_0^{CC}(\uparrow\uparrow)+\frac{1}{2}\xi^{C}L_z   & H_0^{CA}(\uparrow\uparrow)
  & H_0^{CC}(\uparrow\downarrow)+\frac{1}{2}\xi^{C}L_- & H_0^{CA}(\uparrow\downarrow)\\
    H_0^{AC}(\uparrow\uparrow)  &  H_0^{AA}(\uparrow\uparrow)+\frac{1}{2}\xi^{A}L_z
  & H_0^{AC}(\uparrow\downarrow)& H_0^{AA}(\uparrow\downarrow)+\frac{1}{2}\xi^{A}L_-\\
    H_0^{CC}(\downarrow\uparrow)+\frac{1}{2}\xi^{C}L_+  &  H_0^{CA}(\downarrow\uparrow)
  & H_0^{CC}(\downarrow\downarrow)-\frac{1}{2}\xi^{C}L_z&  H_0^{CA}(\downarrow\downarrow)\\
    H_0^{AC}(\downarrow\uparrow)  &  H_0^{AA}(\downarrow\uparrow)+\frac{1}{2}\xi^{A}L_+
  & H_0^{AC}(\downarrow\downarrow)&  H_0^{AA}(\downarrow\downarrow)-\frac{1}{2}\xi^{A}L_z\\
\end{pmatrix}
\quad
\end{align}
\end{widetext}
where $L_{\pm}=L_x\pm i L_y$. $L_{x,y,z}$ are the angular momentum operators.
$\xi^{C}$ ($\xi^{A}$) is the SOC parameter for cation (anion) $p$ orbital.
The typical literature SOC parameters $\xi^{Pb}$=0.91 eV, $\xi^{S}$=0.05 eV,
$\xi^{Se}$=0.22 eV, $\xi^{Te}$=0.49 eV and $\xi^{Po}$=1.06 eV are adopted
from Wittel's spectral data~\cite{wittel1974atomic} in this paper.

\section{Band gap anomaly in P\lowercase{b}X}

\begin{figure}[htp]
\includegraphics[clip, width=3.5in]{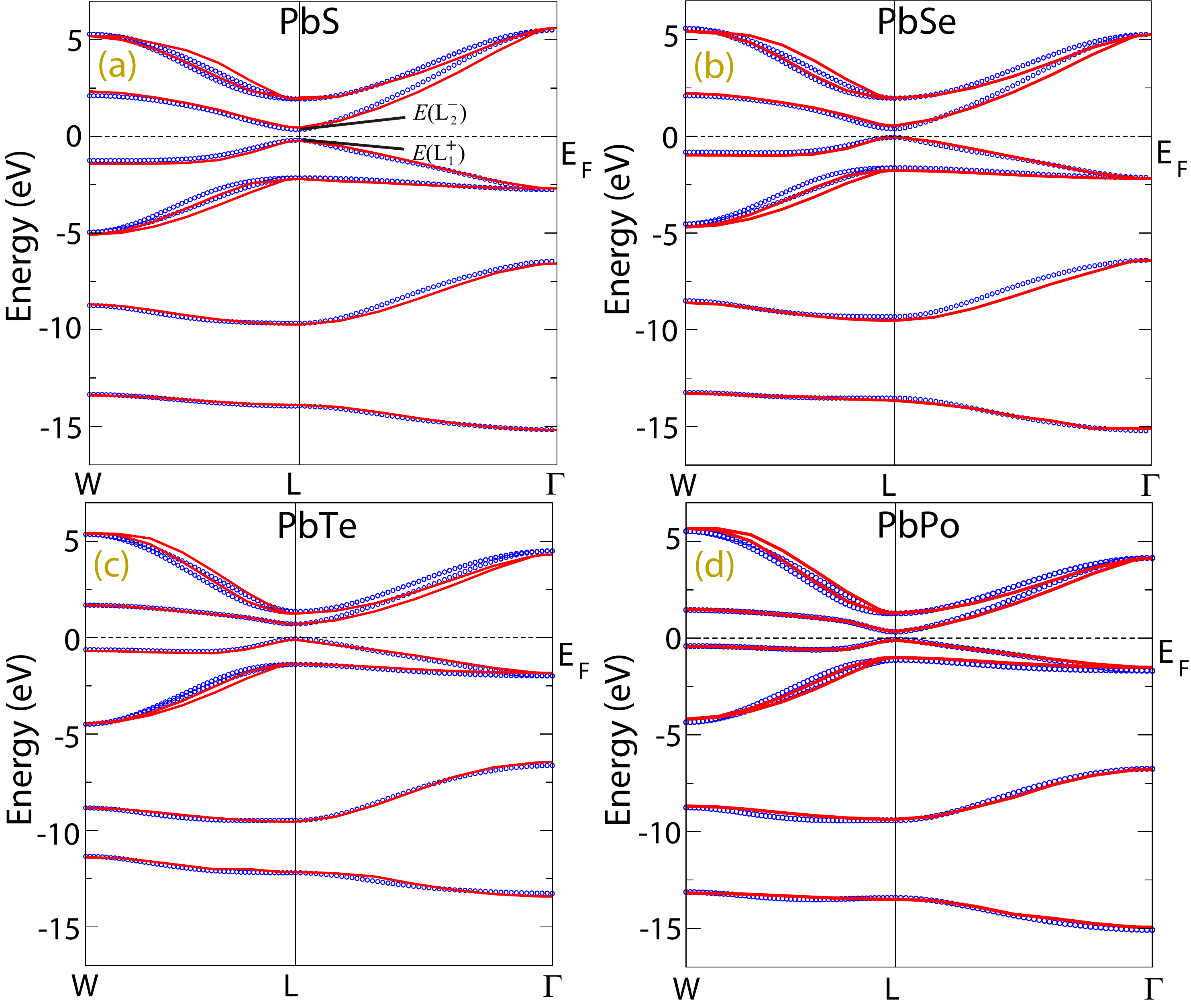}
      \caption{(color online) Non-SOC band structures of PbX. The red lines and the blue
        dots are obtained from HSE03 calculations and TB model fitting respectively.}
\label{fittedbands}
\end{figure}

\begin{figure}[htp]
\includegraphics[clip, width=3.5in]{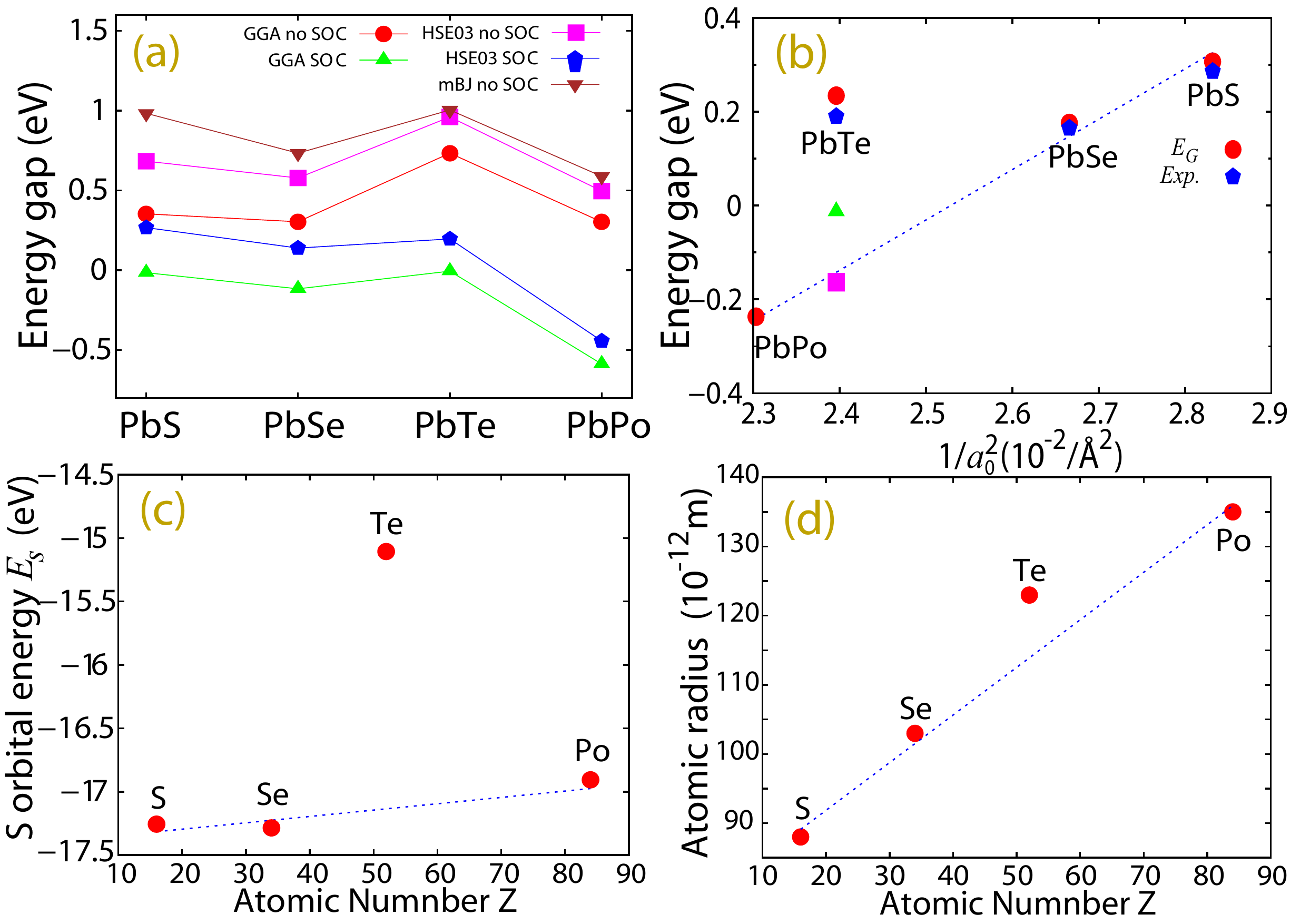}
\caption{(color online) (a) Band gaps $E_G$ at L point for PbX calculated by different methods.
(b) TB band gap $E_G$ for PbX as a function of 1/$a_0^2$.
Available experimental data at 4.2 K (blue pentagons) are plotted for comparison.
(c) and (d) Outermost $s$-electron binding energy $E_s$ and atomic radius for S, Se, Te and Po
as a function of the atomic number Z.}\label{anomal}
\end{figure}

The band gaps calculated by different methods for PbX are shown in Fig. \ref{anomal}(a),
in which the non-SOC band gap is defined as $\Delta E$=$E(L_2^-)$$-$$E(L_1^+)$ as shown in Fig. 2(a).
In Fig. \ref{anomal}(a), the band gaps for PbTe show a distinct jump in all calculation
methods, which means the band gap anomaly in PbTe is intrinsic of the material, and irrelevant to
SOC and the form of the exchange-correlation functional. Comparing all the results shown in
Fig. \ref{anomal}(a), we find HSE03+SOC is the most superior method, which gives almost the same
band gaps as the experimental measurements for all materials. Therefore, in the following, the
band structures calculated by HSE03 without SOC are taken as reference to fit the parameters
in Eq. (\ref{H00}), and all parameters are directly transferred to Eq. (\ref{HH}) by adding the
isotropic SOC term to estimate the band gaps and dispersions for all compounds. The comparisons
between the non-SOC band structures obtained by Eq. (3) and original HSE03 calculations are shown in
Fig. \ref{fittedbands}, and all fitted parameters are listed in Table \ref{fittedparameters}, in which
the detailed definitions of them are given in the six and seven columns. Taking $D_3$ as an example,
$``x,s(100)"$ and $``C-A"$  mean $D_3$ is the hopping parameter from the $p_x$ orbital ($x$) of cation ($C$)
to the $s$ orbital ($s$) of anion ($A$) along the vector $a_0$/2 (1,0,0). In Fig. 2, the non-SOC band
structures obtained from HSE03 calculations and TB fittings by Eq. (3) are plotted with red lines and blue
dots respectively. It is clear that our TB results agree with HSE03 calculations very well, which certifies
that our model is accurate enough to study all details of the system.

\begin{table}
\renewcommand{\arraystretch}{1.0}
\caption{Fitted TB parameters (in eV).}
\begin{tabular}{  l  c  c  c  c  c c c}
\hline
\hline
    &~~ PbS     & ~~   PbSe    &  ~~   PbTe    & ~~  PbPo     &        &       \\
\hline
$A_1$ &~~   -1.1333 & ~~   -2.0072  & ~~   -1.6603  &~~  -2.2198  & ~~  s,s(000)  & ~~  C-C  \\
$A_2$ &~~   ~0.0991 & ~~~   0.0891  & ~~~   0.0959  &~~~  0.0857  & ~~  s,s(110)  & ~~  C-C  \\
$A_3$ &~~   -0.0012 & ~~   -0.0215  & ~~   -0.0406  &~~  -0.0623  & ~~  s,x(110)  & ~~  C-C  \\
$A_6$ &~~   ~7.7801 & ~~~   6.9910  & ~~~   7.2422  &~~~  6.7530  & ~~  x,x(000)  & ~~  C-C  \\
$A_7$ &~~   -0.0002 & ~~   -0.0259  & ~~   -0.0574  &~~  -0.0830  & ~~  x,x(110)  & ~~  C-C  \\
$A_8$ &~~   -0.0169 & ~~   -0.0283  & ~~   -0.0419  &~~  -0.0535  & ~~  x,x(011)  & ~~  C-C  \\
$A_9$ &~~   ~0.3623 & ~~~   0.3110  & ~~~   0.2693  &~~~  0.2157  & ~~  x,y(110)  & ~~  C-C  \\
$C_1$ &~~   -6.9050 & ~~   -7.8493  & ~~   -5.0836  &~~  -7.5326  & ~~  s,s(000)  & ~~  A-A  \\
$C_2$ &~~   -0.0793 & ~~   -0.0848  & ~~   -0.0907  &~~  -0.0894  & ~~  s,s(110)  & ~~  A-A  \\
$C_3$ &~~   ~0.1584 & ~~~   0.1689  & ~~~   0.1788  &~~~  0.1875  & ~~  s,x(110)  & ~~  A-A  \\
$C_6$ &~~   ~3.3582 & ~~~   3.1139  & ~~~   4.2744  &~~~  4.2034  & ~~  x,x(000)  & ~~  A-A  \\
$C_7$ &~~   ~0.1611 & ~~~   0.1686  & ~~~   0.1740  &~~~  0.1810  & ~~  x,x(110)  & ~~  A-A  \\
$C_8$ &~~   -0.0170 & ~~   -0.0188  & ~~   -0.0203  &~~  -0.0217  & ~~  x,x(011)  & ~~  A-A  \\
$C_9$ &~~   ~0.1480 & ~~~   0.1356  & ~~~   0.1201  &~~~  0.1106  & ~~ x,y(110)  & ~~  A-A  \\
$B_1$ &~~   -0.1537 & ~~   -0.1257  & ~~   -0.1082  &~~  -0.0779  & ~~  s,s(200)  & ~~  C-C  \\
$B_2$ &~~   ~0.0855 & ~~~   0.0974  & ~~~   0.1113  &~~~  0.1247  & ~~  s,x(200)  & ~~  C-C  \\
$B_4$ &~~   ~0.5301 & ~~~   0.4891  & ~~~   0.4579  &~~~  0.4131  & ~~  x,x(200)  & ~~  C-C  \\
$B_5$ &~~   ~0.1262 & ~~~   0.1050  & ~~~   0.0870  &~~~  0.0605  & ~~  y,y(200)  & ~~  C-C  \\
$E_1$ &~~   -0.2421 & ~~   -0.2215  & ~~   -0.1947  &~~  -0.1701  & ~~  s,s(200)  & ~~  A-A  \\
$E_2$ &~~   -0.0854 & ~~   -0.1308  & ~~   -0.1733  &~~  -0.2168  & ~~  s,x(200)  & ~~  A-A  \\
$E_4$ &~~   ~0.1210 & ~~~   0.1409  & ~~~   0.1516  &~~~  0.1713  & ~~  x,x(200)  & ~~  A-A  \\
$E_5$ &~~   ~0.0298 & ~~~   0.0330  & ~~~   0.0360  &~~~  0.0393  & ~~  y,y(200)  & ~~  A-A  \\
$D_1$ &~~   -0.6586 & ~~   -0.6076  & ~~   -0.5565  &~~  -0.5015  & ~~  s,s(100)  & ~~  C-A  \\
$D_2$ &~~   ~1.3365 & ~~~   1.2714  & ~~~   1.2389  &~~~  1.1757  & ~~  s,x(100)  & ~~  C-A  \\
$D_3$ &~~   -1.6336 & ~~   -1.4507  & ~~   -1.4907  &~~  -1.2878  & ~~  x,s(100)  & ~~  C-A  \\
$D_4$ &~~   ~1.8308 & ~~~   1.8474  & ~~~   1.8633  &~~~  1.8795  & ~~  x,x(100)  & ~~  C-A  \\
$D_5$ &~~   -0.2662 & ~~   -0.2950  & ~~   -0.3287  &~~  -0.3675  & ~~  y,y(100)  & ~~  C-A  \\
$F_1$ &~~   ~0.3581 & ~~~   0.3100  & ~~~   0.2688  &~~~  0.2239  & ~~  s,s(111)  & ~~  C-A  \\
$F_2$ &~~   -0.0504 & ~~   -0.0491  & ~~   -0.0481  &~~  -0.0470  & ~~  x,s(111)  & ~~  C-A  \\
$F_3$ &~~   -0.0071 & ~~   -0.0069  & ~~   -0.0067  &~~  -0.0065  & ~~  s,x(111)  & ~~  C-A  \\
$F_4$ &~~   ~0.1154 & ~~~   0.1013  & ~~~   0.0868  &~~~  0.0736  & ~~  x,x(111)  & ~~  C-A  \\
$F_5$ &~~   ~0.0315 & ~~~   0.0425  & ~~~   0.0564  &~~~  0.0680  & ~~  x,y(111)  & ~~  C-A  \\
\hline
\hline
\end{tabular}
\label{fittedparameters}
\end{table}

The estimated band gaps $E_G$ (red circles) from Eq. (\ref{HH}) for all four compounds,
as well as the experimental data at 4.2 K (blue pentagons), are summarized in Fig. \ref{anomal}(b).
It is clear that our TB results overlap with the experimental data almost, which certifies
that our model is accurate and powerful again. In order to study the origin of the band gap
anomaly in PbTe, we have checked all the parameters in Table \ref{fittedparameters} carefully
and found that the $s$ orbital onsite energy of Te ($C_1$) is obviously higher than the other
three. Taking $A_1$ (onsite energy of cation $s$ orbital) as a referential energy, we get
$C_1$-$A_1$ equal to -5.7716, -5.8421, -3.4233 and -5.3128 for PbS, PbSe, PbTe and PbPo respectively.
As a result, the band energy of Te $5s$ orbital at $\Gamma$ point shown in Fig. \ref{fittedbands}(c) is
about -13 eV, while it is nearly -15 eV for the other three compounds. Owing to this high onsite energy,
Te $5s$ orbital will push up Pb $6p$ orbital through the $s$-$p$ hybridization and result in a large
band gap for PbTe. Now, we try to artificially decrease the onsite energy of Te $5s$ orbital and push
down the band energy at $\Gamma$ point until it reaches the $``\text{regular}"$ level -15 eV as the
other three. (For this purpose, $C_1$ needs to be decreased by 2 eV.) After this adjustment, though the
band gap does decrease about 0.2 eV as shown as the green triangle in Fig. \ref{anomal}(b), the band gap
anomaly is still evident. Therefore, there must be some other reasons responsible to the band gap anomaly
in addition to the high onsite energy of Te 5s orbital. We then recheck all the parameters in
Table \ref{fittedparameters}, and find that $D_3$ also show some irregularity. Naturally, $|D_3|$ should
decrease monotonically from PbS to PbPo due to the increasing lattice constant. However, $|D_3(\text{PbTe})|$
is a little larger than $|D_3(\text{PbSe})|$, even though $a_0$(PbTe) = 6.460 \AA~ is obviously larger than
$a_0$(PbSe) = 6.124 \AA. Based on the above discussions, after decreasing $C_1$ by 2 eV and increasing $D_3$
by 0.08 eV to a $``\text{regular}"$ number, the final band gap for PbTe is shown as the purple square in
Fig. 3(b), which falls on the line formed by PbS-PbSe-PbPo almost. Therefore, we conclude that the band gap
anomaly in PbTe is mainly related to the high onsite energy of Te 5$s$ orbital and the irregularly large
$s$-$p$ hopping.

Because SOC nearly has no contribution to the band gap anomaly as identified by our $ab ~initio$
calculations, we would like to analytically discuss the band gap evolution and anomaly based on the
non-SOC TB model in the following. At this case, the band gap $\Delta E$ can be easily written as:
{\fontsize{6.8}{4}\selectfont
\begin{align}
\Delta E\!&=\!\frac{1}{2}[A_6\!-\!8A_9\!-\!6E_1-\!4B_5\!-\!2B_4\!+\!C_1\!
\!+\!\sqrt{48 D_3^2\!+\!(A_6\!-\!8A_9\!+\!6E_1-\!4 B_5\!-\!2B_4\!-\!C_1\!)^2} \nonumber\\
-&(A_1\!-\!8C_9\!-\!6B_1\!-\!4E_5\!-\!2E_4\!+\!C_6\!
 +\!\sqrt{48 D_2^2\!+\!(-A_1\!-\!8 C_9\!+\!6B_1\!-\!4E_5-\!2E_4\!+\!C_6\!)^2})] \label{gap}
\end{align}}
which shows an explicit relation between $\Delta E$, $C_1$ and $D_3$.
In Eq. (\ref{gap}), if we omit $D_3$, all $C_1$ ($E_1$) terms will cancel with each other, which
means that the anomaly of $C_1$ needs the help of $D_3$ to involve in the band gap formation and
finally leads to a large gap for PbTe. Therefore, besides the directly bonded $p$-$p$ hopping
parameters ($A_9$, $C_9$, \emph{etc.}), the key factors in determining the band gap for PbX are the
huge $s$-$p$ hybridization $D_3$ ($D_2$) and the irregular high onsite energy $C_1$. We note that,
such huge $s$-$p$ hybridization is a typical character of the covalent systems rather than the ionic
compounds. Generally, the hybridizations between the bonding (antibonding) states and non-bonding
states are very weak in ionic compounds, such as CaX (X=S, Se, Te), SrX (X=S, Se, Te) and BaX (X=S, Se,
 Te)~\cite{dalven1973empirical}, in which the irregular high onsite energy of Te $s$ orbital has
 negligible influence on the band gap $E_G$ because the $s$ (antibonding state from cation)-$s$
(non-bonding state from anion) hybridization is negligible. This is the main reason why CaX, SrX and BaX
all obey the empirical relation $E_G\propto$1/$a_0^2$  very well, even though the irregularity of the
Te $s$ orbital also exists in all these ionic systems. Moreover, we note that the $s$-$p$ hybridizations
$D_3$ ($D_2$) in lead chalcogenides are too strong to be treated by the perturbation method~\cite{Barone2013}.
Using the fitted parameters for PbTe to do a rough estimation, $48D_3^2$ is almost equal to
$(A_6\!-\!8A_9\!+\!6E_1-\!4B_5\!-\!2B_4\!-\!C_1\!)^2$. Therefore, the first square root in the right hand
of Eq. (\ref{gap}) is approximately equal to
$1/\sqrt{2}[\sqrt{48} |D_3|+(A_6\!-\!8A_9\!+\!6E_1-\!4B_5\!-\!2B_4\!-\!C_1)]$, which can be used to
estimate the band gap evolution with $C_1$ and $D_3$ roughly.

Next, we would like to discuss the origin of the irregularity of $C_1$ and $D_3$ deeply.
Because both $C_1$ and $D_3$ are related to the $s$ orbital of anion tightly, it is
natural to propose that the irregularity may be originated from the chalcogen atoms themselves.
We have calculated the binding energies of the outermost $s$ electrons $E_s$ for all chalcogen
atoms and plotted them with the atomic number Z in Fig. \ref{anomal}(c), in which the points
for sulfur, selenium and polonium show an approximately straight line while the point of tellurium
is much higher than the straight line. This result clearly demonstrates that the high onsite energy
$C_1$ of PbTe is inherited from the low binding energy of tellurium $5s$ electrons. To study the
irregularity of $D_3$ in PbTe, we show a plot of the atomic radius of the chalcogen
atoms~\cite{clementi1967atomic} as a function of atomic number Z in Fig. \ref{anomal}(d).
We can see that the atomic radius of tellurium does not lie on the straight line as the others do and
its position is higher than the line, which means tellurium $5s$ orbital is more extended. As a result,
the hybridization between $5s$ orbital of Te and $p_x$ orbital of Pb, \emph{i.e.} $D_3$, is abnormally
larger than the $``\text{regular}"$  value. Both the low binding energy and the extended distribution
of the tellurium $5s$ electrons are related to the unusually strong screening effect of the V-period
elements, as well as the penetration effect and relativistic effect. In general, the huge screening effect
from the interior electrons will reduce the binding effect of nucleus and lead to a low binding energy
and more extended distribution of Te 5s electrons. However, the detailed discussions of the atomic energy
level and distribution are very complicated and out of the scope of this work. Finally, we claim that
such $5s$ electrons irregularity is universal for the right side elements in the V row of the periodic Table of
Elements \cite{sevier1979atomic}, \emph{e.g.} Sn, and our discussion is universal for other covalent
systems with the same structure. For example, similar analysis would give the conclusion that the $5s$
orbital irregularity of Sn will obviously reduce the band gap of the tin chalcogenides, which may be the
main reason why the band gap of tin chalcogenide is usually smaller than the same row lead
chalcogenide\cite{sun2013,Tsuoptical}.

\section{TOPOLOGICAL PROPERTIES IN P\lowercase{b}P\lowercase{o}}

\begin{figure}[tbp]
\includegraphics[clip, width=3.5in]{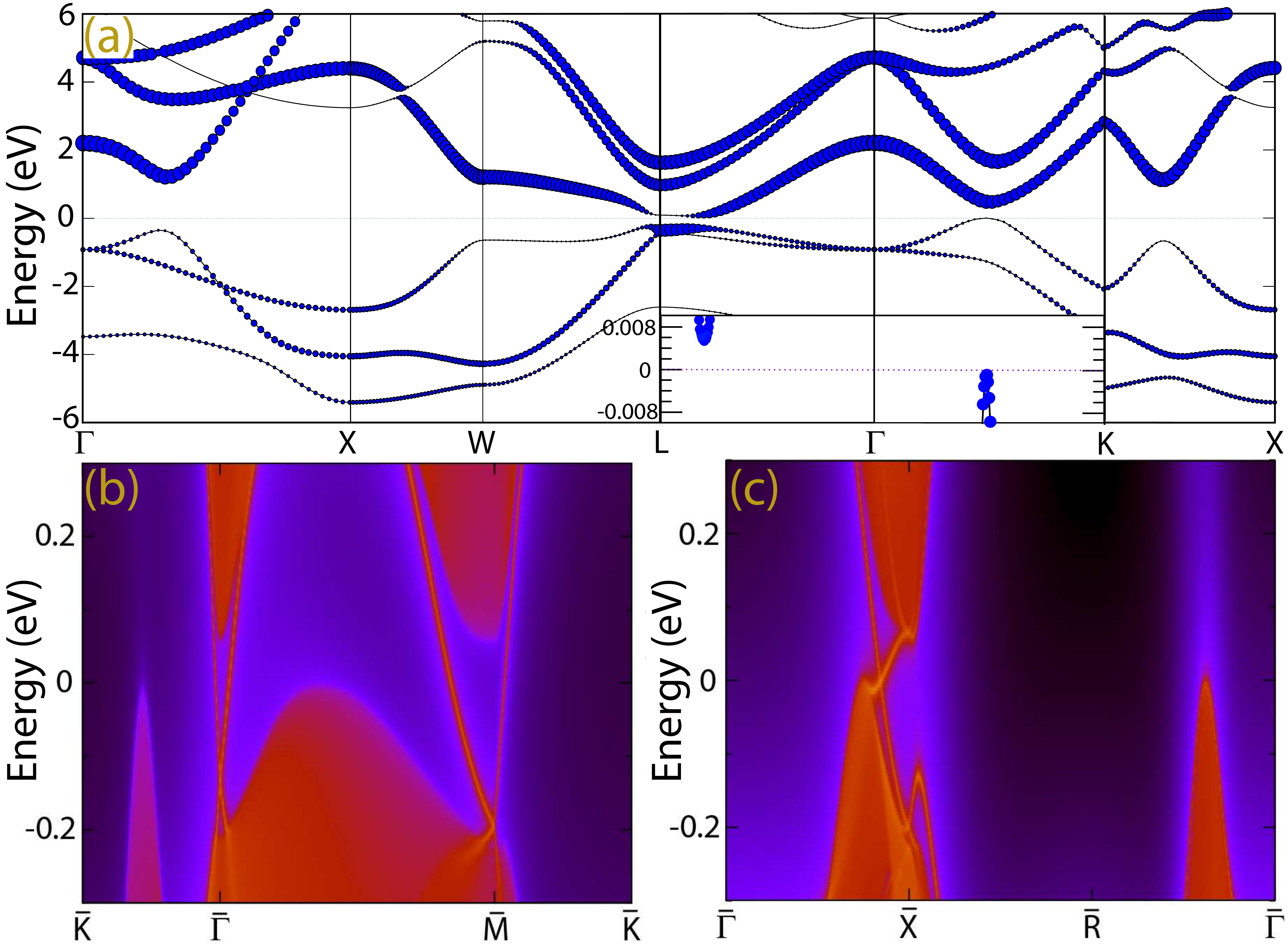}
      \caption{(color online) (a) Band structures of PbPo calculated by HSE03+SOC. The size of the blue circles represents the
       weight of the projected $p$ orbitals of Pb. The inset shows the top of the valence band and the bottom of the conduction
       band of PbPo. (b) and (c) The calculated surface states on (111) and (001) planes.}
\label{surface}
\end{figure}

One important result of our calculations is that $E_G$ is negative for PbPo as shown in
Fig. \ref{anomal}(b), which means that band inversion happens at L point of PbPo. Detailed
HSE03+SOC calculations have been performed to confirm this conclusion. The calculated band
structure is shown in Fig. \ref{surface}(a), in which one $p$ orbital of Pb is obviously
dropped down below the Fermi level at L point as represented by the blue circles. Furthermore,
our calculations show that PbPo is an indirect band gap (6.5 meV) semiconductor (see inset
of Fig. \ref{surface}(a)), rather than a direct band gap semiconductor\cite{dalven1972} or
semimetal\cite{Rabii1974}. All these band characters of PbPo are very similar to SnTe,
which implies that PbPo may be a TCI too.

There are four L points  \{L$_1$ (0, $\pi$, 0), L$_2$ (0, 0, $\pi$), L$_3$ ($\pi$, $\pi$, $\pi$),
L$_4$ ($\pi$, 0, 0)\} in the first BZ of PbPo (see Fig. \ref{structure}(b)). Because there are
even band inversion points, the topological property of PbPo is different with 3D strong TI~\cite{zhang2009,zhang2011topological}.
Actually, the topological property is protected by the mirror symmetry, and its topological invariant
is the mirror Chern number $C_M$ instead of $Z_2$.

The eigenvalue ($m$) of the mirror operator $\hat{m}_{(1\overline{1}0)}$ is a good quantum number for PbPo.
So we can classify the Bloch wavefunctions on (1$\overline{1}$0) plane by $m$, and define the berry
connection $\mathbf{A}^m(\mathbf{k})$ and berry curvature $\Omega^m(\bf{k})$ on the plane as follows:
\begin{align}
\mathbf{A}^m(\mathbf{k})&=  i\sum_n \langle u^m_n(\mathbf{k})|\nabla_{\mathbf{k}}|u^m_n(\mathbf{k})\rangle \\
\Omega^m(\mathbf{k})&=\nabla_{\mathbf{k}}\times\mathbf{A}^m(\mathbf{k})
\end{align}%}
where $u^m_n(\mathbf{k})$ with mirror eigenvalue $m$ = $\pm i$ is the $n$th eigenstate at
$\mathbf{k}$ point. The sum is over all occupied bands. The mirror Chern number $C_M$ is
defined as $C_M=(n_{+i}-n_{-i})/2$, where $n_{m}=\int \Omega^m(\mathbf{k}) \cdot d\mathbf{S}$.
Using above formula, we find $C_M=-2$ for PbPo, which confirms that PbPo is a TCI.
The other hallmark of the TCI is the mirror symmetry protected nontrivial surface states.
We have calculated the surface states on (111) and (001) plane as shown in Fig. \ref{surface}(b) and
Fig. \ref{surface}(c), respectively. As we can see, there are two distinct Dirac cones:
one is pinned at the time reversal invariant momentum (TRIM) point while the other is situated
off the TRIM point. On (111) plane, L$_2$, L$_3$ and L$_4$ are projected to the same point
$\overline{\text{M}}$, while L$_1$ is projected to $\overline{\Gamma}$ point. There is no additional
interaction coming from the scattering between odd L-valleys. So the Dirac cones on (111) plane
exactly locate at the TRIM points as shown in Fig. 4(b). On (001) plane, L$_1$ and L$_2$ ( L$_3$ and L$_4$)
are projected to the same point $\overline{\text{X}}$ (another $\overline{\text{X}}$). The interaction
between two L-valleys will introduce an additional term to push the Dirac cones away from TRIM
point~\cite{liu2013}. However, k-points along $\overline{\Gamma}-\overline{\text{X}}$ direction preserve
mirror symmetry with respect $\hat{m}_{(1\overline{1}0)}$ operator. Therefore, even the Dirac cones are
pushed away from TRIM point, they can still survive along $\overline{\Gamma}-\overline{\text{X}}$
direction as shown in Fig. 4(c). The Dirac cones are protected by the mirror symmetry rather than TR
symmetry, which is an important difference between TCI and $Z_2$ TI.

\section{CONCLUSION}

In summary, we have studied the band evolution in PbX, and found that, though the $s$ orbital
of the anion is far away from the Fermi level, it has crucial influence on the band gap through
the huge $s$-$p$ hybridization. The high onsite energy of Te $s$ orbital and its
irregular extended distribution combining together result in an irregular big band gap in PbTe,
\emph{i.e.} the famous band gap anomaly in lead chalcogenides. Furthermore, our calculations show
that PbPo is an indirect band gap (6.5 meV) semiconductor with negative $E_G$, which means that
band inversion happens at L point in PbPo. The calculated mirror Chern number and
surface states confirm that PbPo is a TCI.

\begin{acknowledgments}

We thank Xianxin Wu, Zhijun Wang, Richard Martin and Theodore Geballe for valuable discussions,
the support from National Science Foundation of China (Grant Nos. 11204359), the 973 program 
of China (Grant Nos. 2013CB921700), and the ¡°Strategic Priority Research Program (B)¡± of the 
Chinese Academy of Sciences (Grant Nos. XDB07020100).
\end{acknowledgments}

\bibliography{bibtex}

\begin{appendix}

\section*{APPENDIX}

In the APPENDIX, the detailed descriptions of the real and imaginary
matrix elements in Eq. (\ref{H00}) are shown in Table \ref{H0} and
Table \ref{G0} respectively.

\begin{widetext}

\begin{table}[htp]
\renewcommand{\arraystretch}{0.8}
\caption{Real part of $H_0(\mathbf{k})$}
\begin{tabular}{  l  l }
\hline
\hline
 $H_{1,1}$  &~~~$A_1+4A_{2}[cos(k_x)cos(k_y)+cos(k_y)cos(k_z)+cos(k_z)cos(k_x)]+2B_{1}[cos(2k_x)+cos(2k_y)+cos(2k_z)]$\\
 $H_{1,5}$  &~~~$2D_1[cos(k_x)+cos(k_y)+cos(k_z)]+8F_1cos(k_x)cos(k_y)cos(k_z)$\\
 $H_{2,2}$  &~~~$A_6+4A_7cos(k_x)[cos(k_y)+cos(k_z)]+4A_8cos(k_y)cos(k_z)+2B_4cos(2k_x)+2B_5[cos(2k_y)+cos(2k_z)]$\\
 $H_{2,3}$  &~~~$-4A_9sin(k_x)sin(k_y)$\\
 $H_{2,4}$  &~~~$-4A_9sin(k_x)sin(k_z)$\\
 $H_{2,6}$  &~~~$2D_4cos(k_x)+2D_5[cos(k_y)+cos(k_z)]+8F_4cos(k_x)cos(k_y)cos(k_z)$\\
 $H_{2,7}$  &~~~$-8F_5sin(k_x)sin(k_y)cos(k_z)$\\
 $H_{2,8}$  &~~~$-8F_5sin(k_x)sin(k_z)cos(k_y)$\\
 $H_{3,3}$  &~~~$A_6+4A_7cos(k_y)[cos(k_z)+cos(k_x)]+4A_8cos(k_z)cos(k_x)+2B_4cos(2k_y)+2B_5[cos(2k_z)+cos(2k_x)]$\\
 $H_{3,4}$  &~~~$-4A_9sin(k_y)sin(k_z)$\\
 $H_{3,6}$  &~~~$H_{2,7}$\\
 $H_{3,7}$  &~~~$2D_4cos(k_y)+2D_5[cos(k_z)+cos(k_x)]+8F_4cos(k_x)cos(k_y)cos(k_z)$\\
 $H_{3,8}$  &~~~$-8F_5sin(k_y)sin(k_z)cos(k_x)$\\
 $H_{4,4}$  &~~~$A_6+4A_7cos(k_z)[cos(k_x)+cos(k_y)]+4A_8cos(k_x)cos(k_y)+2B_4cos(2k_z)+2B_5[cos(2k_x)+cos(2k_y)]$\\
 $H_{4,6}$  &~~~$H_{2,8}$\\
 $H_{4,7}$  &~~~$H_{3,8}$\\
 $H_{4,8}$  &~~~$2D_4cos(k_z)+2D_5[cos(k_y)+cos(k_x)]+8F_4cos(k_x)cos(k_y)cos(k_z)$\\
 $H_{5,5}$  &~~~$C_1+4C_2[cos(k_x)cos(k_y)+cos(k_y)cos(k_z)+cos(k_z)cos(k_x)]+2E_1[cos(2k_x)+cos(2k_y)+cos(2k_z)]$\\
 $H_{6,6}$  &~~~$C_6+4C_7cos(k_x)[cos(k_y)+cos(k_z)]+4C_8cos(k_y)cos(k_z)+2E_4cos(2k_x)+2E_5[cos(2k_y)+cos(2k_z)]$\\
 $H_{6,7}$  &~~~$-4C_9sin(k_x)sin(k_y)$\\
 $H_{6,8}$  &~~~$-4C_9sin(k_x)sin(k_z)$\\
 $H_{7,7}$  &~~~$C_6+4C_7cos(k_y)[cos(k_z)+cos(k_x)]+4C_8cos(k_z)cos(k_x)+2E_4cos(2k_y)+2E_5[cos(2k_z)+cos(2k_x)]$\\
 $H_{7,8}$  &~~~$-4C_9sin(k_y)sin(k_z)$\\
 $H_{8,8}$  &~~~$C_6+4C_7cos(k_z)[cos(k_x)+cos(k_y)]+4C_8cos(k_x)cos(k_y)+2E_4cos(2k_z)+2E_5[cos(2k_x)+cos(2k_y)]$\\
\hline
\hline
\end{tabular}
\label{H0}
\end{table}

\end{widetext}

 %\begin{widetext}

\begin{table}[htp]
\renewcommand{\arraystretch}{0.7}
\caption{Image part of $H_0(\mathbf{k})$}
\begin{tabular}{  l  l }
\hline
\hline
  $G_{1,2}$ &~~~$-4A_3sin(k_x)[cos(k_y)+cos(k_z)]-2B_2sin(2k_x)$\\
  $G_{1,3}$ &~~~$-4A_3sin(k_y)[cos(k_z)+cos(k_x)]-2B_2sin(2k_y)$\\
  $G_{1,4}$ &~~~$-4A_3sin(k_z)[cos(k_x)+cos(k_y)]-2B_2sin(2k_z)$\\
  $G_{1,6}$ &~~~$ 2D_2sin(k_x)+8F_3sin(k_x)cos(k_y)cos(k_z)$\\
  $G_{1,7}$ &~~~$ 2D_2sin(k_y)+8F_3sin(k_y)cos(k_z)cos(k_x)$\\
  $G_{1,8}$ &~~~$ 2D_2sin(k_z)+8F_3sin(k_z)cos(k_x)cos(k_y)$\\
  $G_{2,5}$ &~~~$-2D_3sin(k_x)-8F_2sin(k_x)cos(k_y)cos(k_z)$\\
  $G_{3,5}$ &~~~$-2D_3sin(k_y)-8F_2sin(k_y)cos(k_z)cos(k_x)$\\
  $G_{4,5}$ &~~~$-2D_3sin(k_z)-8F_2sin(k_z)cos(k_x)cos(k_y)$\\
  $G_{5,6}$ &~~~$-4C_3sin(k_x)[cos(k_y)+cos(k_z)]-2E_2sin(2k_x)$\\
  $G_{5,7}$ &~~~$-4C_3sin(k_y)[cos(k_z)+cos(k_x)]-2E_2sin(2k_y)$\\
  $G_{5,8}$ &~~~$-4C_3sin(k_z)[cos(k_x)+cos(k_y)]-2E_2sin(2k_z)$\\
\hline
\hline
\end{tabular}
\label{G0}
\end{table}

\end{appendix}

 \end{document}